**Michael E. Fisher – teacher, mentor, colleague and friend: a (very) personal account.**

Eytan Domany

Department of Physics of Complex Systems
Weizmann Institute of Science, Rehovot 7630031, Israel
eytan.domany@weizmann.ac.il

*"The only rational way of educating is to be an example. If one can't help it, a warning example."*                                        Albert Einstein.

I had the good fortune and privilege of having Michael Fisher as my teacher, supervisor, mentor and friend. During my years as a scientist, teacher and supervisor of about one hundred students and post docs I found myself innumerable times realizing that I am following or at least trying to follow Michael's example. These pages are my attempt to convey recollections of my association with Michael, focusing on how he served as an example for me.

1. Introduction

The most difficult part of writing a memorial for someone close to you is to get started – to decide how to go about it. The natural way is to put things in temporal order, but perhaps a more attractive alternative is a random walk in time and space, allowing associations to dictate the sequence in which events come to mind. The second aspect in terms of difficulty is to keep in mind that you are not writing about yourself, but about Michael, and this is really hard, since the recollections you record are personal and naturally you are a central player in whatever you reminiscence about.

I decided to be flexible on both accounts – while trying to keep chronological order, allowing my mind to take leaps when they pop up, and on the personal front – to shrug off sanctimonious reservations and criticism of self-promotion, and talk about myself whenever I feel like it.

2. Pre Ph D, first year at Cornell

My first encounter with Fisher was in early 1972. I was working at a research center in Israel, finishing my M Sc. After lunch we used to pass through the library and I remember distinctly one of my older colleagues (he was probably about half my current age) picking up the issue of the Physical Review Letters that just arrived, opening it and exclaiming: there is really no limit to what people of leisure will publish! He then read aloud the title of a Letter: Critical Exponents in 3.99 Dimensions, by Wilson and Fisher.[1] I was intrigued, and took a course on critical phenomena that briefly mentioned the Renormalization Group.

For a variety of personal reasons I decided, some months later, to do a Ph D in the US. My M Sc advisor, Michael Kirson, recommended to me Cornell, where he did his Ph D with Hans Bethe. I applied, was accepted and we (Pnina, my recently married wife and I) found ourselves in Ithaca in late August 1973. I was immediately immersed in the small community of Israeli graduate students, mainly from Business Administration, Political Science and the like, who were in fierce competition to get their degrees as fast as possible, so during the first semester I took an unreasonably heavy workload of courses. One of these was an advanced Statistical Physics seminar given by Geoffrey Chester, in which students were supposed to propose and complete some research project. When the time came I mentioned that I heard that something interesting was happening in critical phenomena, was curious to find out what it was, and that I would also like to do a Monte Carlo simulation. Chester said: if so, let's go and talk to Fisher![a] So a meeting was set up, and Chester, K.K. Mon (my partner in the project) and I walked over from Clark Hall to see Fisher in his office in Baker.

Michael was very kind, friendly and straight. He told us that he believed that a good Monte Carlo was a dead Monte Carlo, but if we insisted, he was willing to give the idea a try. He agreed to propose a question, which on the one hand he did find interesting, but on the other he had a good idea what the answer should be, so that if we generated complete nonsense he would be able to tell. The problem he suggested was simulations of the 8-vertex model, solved exactly by Baxter[2], which appeared as a special "multicritical" case in a much broader space of models he wanted to study. I dwell on this project and the resulting publication[3] because it highlights several attributes of Michael, which had a profound impact on my own approach and attitude to science, lasting throughout my career.

***Michael's dead serious attitude to his science*** *and research was my first striking take-away lesson.* Once he set out on a problem, even if it was a mere course project of two first year graduate students, he treated it with full, uncompromising attention to every detail. The problem was embedded in the most general setting, the research questions were clearly and precisely stated, no stone was left unturned and no corners were ever cut. As part of our project, background material was thoroughly read and reviewed – K.K. and I read papers on the Baxter model,[2] its spin representation,[4] marginal operators, scaling hypothesis[4]; we learned finite size scaling theory[5] and how to do Monte Carlo simulations.[6] We started by simulating Ising models, to demonstrate that we did it right, deriving known results from finite size scaling, and only then attacking the full problem.[b]

Michael met us and reviewed our progress periodically, ever suspicious, insisted on testing and retesting every finding. When he was satisfied with the results – agreement of the numerically derived continuous variation of the specific heat exponent on the Baxter line with the known answer, confirmation of the various scaling hypotheses and the effect of symmetry breaking, he told me that I should now write a first draft of a paper.

---

[a] Forty years later Michael wrote me that he first met Chester when he started graduate work in 1952. In 1974 I was unaware of their long standing connection.

[b] Our computer program was punched onto cards that we took down to the basement of Clark, and fed them into a card reader connected to a CDC 7600 machine at Berkeley. At late hours our only competitor at the reader was Ken Wilson, who ran his Exact Renormalization Group programs on the same machine. Between runs we competed with Wilson also on a computer game, landing a simulator of a moon landing on a small green screen, which was part of a huge computer that controlled low temperature experiments of Richardson, Lee and Reppy.

I was quite literate in English and took pride in my writing skills, and hence was stunned when my attempt came back completely rewritten in red ink between my lines. There was, however, a comment on top: "Not bad at all for a first draft!" followed by an immortal Fisher line: "but why do you use the *blunt end of a charred broomstick* to draw your figures?"

**Michael's approach to writing his papers** *was my next lesson.* The student who worked on the project always wrote the first draft, as part of his training, even if rewriting it was far more time consuming than just writing it. Michael revised our paper several times, devoting to it immense attention and energy. Since his name was on the paper, it had to meet his standards of clarity, scientific and linguistic precision, and style. The number of future readers of the paper is completely irrelevant; there is only one person whose standards and quality control matter, and that person is Michael, and for all my future papers – me. This statement holds for the present manuscript too!

## 3. Ph D – first year

At the end of the summer of 1974, having passed my qualifying exam, I went to see Michael to check the possibility of doing a Ph D under his supervision. "Going to see him" calls for some explaining about the setup at Baker.

The main entry to Michael's office was via a door that led to a tiny entrance space. To the right there was a windowless cubicle where his Senior Graduate Student, David Nelson, sat. Straight ahead was a room occupied by Lynn, the secretary he shared with Ben Widom. From this room opened the doors to Ben's office on the left and Michael's on the right. You asked Lynn to set up an appointment with Michael, and she told you over the phone the day and time. When the front door of this complex was locked (e.g. after office hours) – you entered a tiny dark space directly from the corridor, and waited there till Michael opened a door to his office. In this small space there were several black and white photographs of some fluid in motion – taken presumably by Michael when he did his Ph D thesis, solving some differential equations. It was through this anteroom that I came to see him. The prime exhibit in his office was a photograph of young Michael, confident and self-assured, facing Lars Onsager.

As always, **Michael was direct, honest and to the point.** He told me that I had to wait since he took only one Ph D student at a time, there was another enquiry on line, and the other candidate was preferred. It was Mike Peskin, a fellow student and friend, who asked also Ken Wilson about a Ph D. When some time later Ken agreed to be his advisor, I got a call to come back and see Michael, who informed me that I was accepted, but first he wanted to discuss with me two issues.

The first, he said with a measure of unease, was that "Israelis were known to be brusque" and he hoped that this will not be the case with me. To which I responded: Professor Fisher, would you mind explaining to me what is "brusque"? which he did, and I assured him that I will be the last person his secretary will ever complain about. I did not enquire what gave him the notion that Israelis were brusque – he had an Israeli postdoc at the time, one Amnon Aharony, who was some 6 years earlier the Teaching Assistant of my undergraduate quantum mechanics course at Tel Aviv University and became a friend and collaborator years after Cornell[c]. Amnon just finished two years with Michael, during which he wrote more than 30 papers, and I could imagine the stress he caused among the secretaries who had to type them (on typewriters, mind you!).

The second issue Michael brought up also caused him some discomfort. At Cornell, once you started working on your Ph D, your advisor paid from a grant your tuition, fees and Fellowship as a Research Assistant. As an RA, you worked full time on your thesis research and did not have to spend about 10 hours a week as a TA, teaching or assisting in a lab.[d] Michael informed me that his current grants did not have funds for an RA and I will have to continue as a TA. Now it was my turn to be embarrassed; I apologized for not bringing this up before, but I did intend to ask him whether it would be possible to work with him without taking a position as his RA. He was very surprised, and I explained that a few months before I found a summer job as RA of Professor Jim Krumhansl. I calculated effective elastic constants of some composite materials, and following the request of Jim Gubernatis, the postdoc in charge of the project, Krumhansl asked me to stay on as RA on a huge DARPA grant he just landed (on ultrasonic detection of defects in the engine of the just approved B1 bomber aircraft). He emphasized that this was "a bread and butter job", i.e. with no Ph D prospect down the line, and he expected me to look for someone else as my thesis advisor. I found out many years later that Michael was profoundly surprised and deeply impressed by my willingness and daring to do a Ph D with him and work in parallel on some other research project. When a year later Michael told me that now he did have funding and I could become his RA, I again asked him if it would be ok with him if I stayed on with Jim Krumhansl? Gubernatis (who became a close friend) got a position at Los Alamos and was leaving Cornell, a new postdoc was expected to arrive several months later and would need time to get into the project, so that Jim really depended on me to run his huge DARPA grant. Michael of course agreed and I mention all this for two reasons: to show **how open minded and non-possessive Michael was**, and to sketch the special and unique atmosphere at Cornell at the time. The advanced students and post docs talked to one another all the time, small ad hoc teams continuously formed and dissolved discussing problems, writing papers. Post docs in particular were roaming freely around. People like Mike Kosterlitz, Alistair Bruce, the Combescots, Mike Peskin, Serge Aubry, Alan Bishop, Steve Teitel and many others were enthusiastically discussing whatever excited them at any moment. I never encountered any barrier due to "I am working with Prof. X and you with Y"; on the contrary, you had to make an effort to filter out stuff that was flowing around you.

So in Fall 1974 I started my Ph D with Fisher. This statement also needs some qualifying; actually I started working with David Nelson. Now I am fully aware that David is alive and well, and it is not his obituary that I am writing, but it is impossible to describe my years at Cornell with Michael without talking about David. He was some years younger than I, on a six-year Ph D program. Clearly one of the sharpest and smartest guys around, he already published seminal work using the epsilon expansion to analyze critical behavior near multicritical points; he just wrote a paper with Rudnick[7] on calculating free energy and tricritical crossover scaling functions by integrating various quantities along RG trajectories. He proposed to Michael using trajectory integrals to calculate crossover scaling functions near bi-critical points, and I was assigned to work on this problem. To say that working with David was challenging is a severe understatement, but we did get along well and remained friends. Some years ago I visited him at Harvard and gave a seminar – he introduced me saying that we were graduate students together and we both had to put up with Fisher, to which I responded – you had to put up with Fisher, I had to put up with Fisher and with you! He was filling spiral notebooks with error-free analytic calculations extremely fast, with his writing speed being the rate-limiting factor. Our project progressed well and we calculated the scaling functions in the disordered phase, with relatively little involvement of Michael, who was not a co-author on this paper[8] (first of a series of three).

---

[c] To do justice to Amnon: at some later time I did ask Michael about this and he mentioned two other very prominent Israeli scientists who passed through Cornell before Amnon. Michael was well aware of the dangers of inference based on a single observation.

[d] Actually, as an RA you were expected to contribute to teaching as a grader of some course. I had the good luck to grade the last Quantum Mechanics course given by Hans Bethe prior to his retirement. It was one of my most amazing experiences at Cornell.

*This was another important lesson I learned from Michael: **let your students initiate projects, and work on them independently, teaming up with other students. Never "sign" papers to which you did not make some significant contribution**.*[e] While at Cornell, I wrote papers alone and with several collaborators. Michael was always supportive but never ever made me feel that as my thesis advisor he should be a co-author. This does not imply that he was indifferent, on the contrary: his door and mind were open to ideas. His curiosity was boundless and if you came with an idea which he found interesting and challenging his eyes would glitter; at times he would get involved and work on it enthusiastically (see an example below). He always found the right balance between letting you work at your own pace, shifting to the problems that excited you most, and acting, at the same time, as advisor and collaborator.

One aspect of Michael's attitude to science and students, which to my regret I did not adopt, was **attending and taking students to conferences**. He went to every major meeting, many times giving keynote invited talks, but when he was interested in learning something new (e.g. biology), he had no ego problem to register and give a poster presentation. He took me along to the major conferences in the field; the first one, which I remember most vividly was a Yeshiva meeting in New York in December 1974. I even gave a 1-minute "talk" standing in the back of a large dark auditorium and screaming something, with the formidable Joel Lebowitz on the stage, shutting up everyone who went over time just by standing up and looking at the speaker with a big smile, without saying a word.

4. **Ph D – second year**

When after my first year David Nelson left for Harvard, Michael and I switched to a closer mode of collaboration. I used to see him on Tuesdays at the "Fisher seminar" (see below) and every Friday for one hour alone in his office. Around that time he became Chair of the department of

---

[e] This got me into trouble with my biologist colleagues: several times a paper I have never seen landed in my inbox with my name on it as a co-author (mostly because one of my students helped with some analysis). When I removed my name, my collaborator was offended – is this paper not good enough for you?

Chemistry, so he was very busy, but my allocated hour was always there. When I did not make much progress and did not have anything of substance to discuss and report, I would call and say so and we cancelled our meeting. At a Magnetism conference to which we went together we met two guys from IBM Zurich, Schneider and Stoll, who were considering sending their student, Stefan Sarbach, to Cornell on a Swiss post-doctoral Fellowship. Michael warned them that he was very busy and advised them to talk to me to find out how was it working with him. I told them that I had Michael's full attention for an hour every Friday, and I recall Toni Schneider asking – is one hour a week with your advisor enough? I could not help reverting to brusque Israeli behavior and told him that it was far better to spend one hour a week with a wise man than 8 hours a day with a fool. I am not sure whether it was this deep insight that convinced them, but Sarbach did come to Cornell. Another person who came, around the same time (Fall 1975), was an old friend of mine, David Mukamel – he and I shared an apartment and a boss in the late sixties. David finished a postdoc at Brookhaven and came for a second one at Cornell. During my next (final) year, while finishing the second[9] and third[10] papers on scaling functions, I collaborated with David and Michael on some other problems, *which serve as the example promised above, of* **Michael being "turned on" by an idea to which he got exposed.**

While at Brookhaven, David identified physical systems whose phase transitions were characterized by order parameters with $n \geq 4$ components.[11] Using Landau theory he wrote down their Landau-Ginzburg Hamiltonians,[f] and using the epsilon expansion, discovered that some of them did not have stable fixed points of the RG.[12] Since the corresponding physical systems had first order transitions, he proposed that lack of stable RG fixed points indicates that a transition of the corresponding system must be first order. This was an extension of the symmetry-based Landau rules[13] that excluded symmetry breaking via a continuous transition.

Michael did not like Landau theory: he viewed it as an embellishment of Mean Field theory by fancy group-theoretical terminology (in which he was not well versed and hence suspicious of its relevance). He did take deep interest, however, in first order transitions. In his course on phase transitions and critical phenomena, which he gave in late 1974 to a huge class of 120 students, Michael emphasized that we do not understand first order transitions. I dared raise my hand and mentioned that there were elaborate Landau rules based on symmetry that excluded the existence of continuous transitions, and hence identified situations in which symmetry breaking must be first order.[13] He gave me a look and said: "the trouble with Israelis is that they take the Russians too seriously!" 119 students burst out laughing.

David did succeed to convince Michael that the Landau framework, based on symmetry, was the natural one to construct correct Landau-Ginzburg Hamiltonians for physical systems undergoing a phase transition. Whatever you do next, e.g. how you approximate the ensuing functional integral, is your business – be it Mean Field, Monte Carlo or your favorite form of RG.

Interest in first order transitions fueled two studies. The first one addressed the question of what happens when a field is applied to a system for which the transition is first order, due either to the absence of a stable fixed point or it being inaccessible under the RG flows. In some cases the field can make a stable fixed point accessible, and we investigated how the transition turns from first order to continuous.[14]

---

[f] I always wondered why was Wilson's name added to this concept?

I was intrigued by the fact that if you look at a cube down the (111) axis, it has 3-fold symmetry. David and I thought about an ordered cubic ferromagnet with the six (100) directions being the easy axes. In a strong magnetic field pointing along (1,1,1) the spins will be aligned with the field. As the field is reduced, the magnetization will flip towards one of three equivalent directions. David showed that this transition is in the 3-state Potts universality class; this system provided a physical realization of the 3-state Potts model in *d*=3 dimensions! There was some evidence indicating that the transition was first order – whereas in *d*=2 it was known to be continuous.[15] Michael got immersed in the problem – it was generalized to external fields in an arbitrary direction, the full phase diagram was mapped out, in which the Potts transition was a special multicritical point. A quick Letter was written,[16] which turned the spotlight to yet another specialty of Michael: *his love of drawing and his talent at it.* The published phase diagrams, for first order and continuous Potts transitions, are remarkable examples of combining art and aesthetics in scientific drawings. Many years later, in our apartment in Rehovot, my 6-year old son Gil was fascinated by Michael's drawings of nudes – in particular, how he was able to bring out their three-dimensional curvature? Michael, ever willing to explain and teach, spent hours showing Gil how to draw female bosoms.

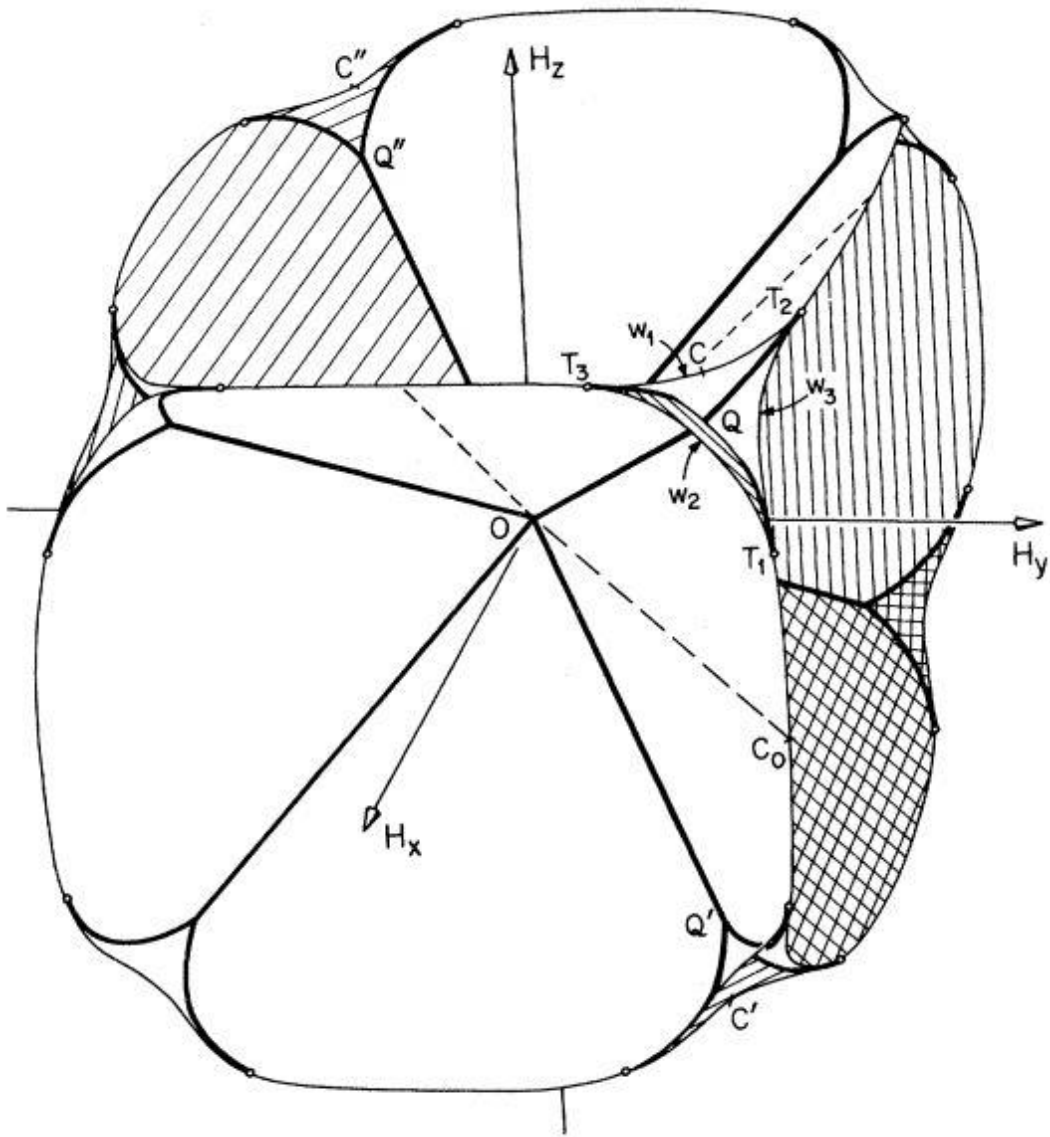

Figure 1. Art in Science: Schematic mean field phase diagram of a cubic ferromagnet in an external field. The 3-state Potts transitions are at point Q (and at symmetry-related points Q', Q''). See reference 16.

**5. Fisher Seminar, teaching, super-referee**

I remember admiring the drawings he made on Styrofoam cups while listening to seminars. The only "formal" requirement that came with being associated with Michael was attending the "Fisher seminar" every Tuesday afternoon. The audience sat around tables; blackboard and chalks were provided to speakers, but transparencies were strictly forbidden.[g] If the speaker wanted to show figures (which would take too long to draw on the blackboard), a large number of copies were made and distributed to the audience. Michael felt that transparencies convey the illusion that whatever was on them was correct, while text written on the blackboard was more inviting discussion and criticism. Besides, writing on blackboard slowed the speaker down, giving the participants time to think and to take notes. Michael was notorious for mercilessly grilling speakers with his questions. The foremost reason was his genuine boundless intellectual curiosity; if he found the topic interesting enough to have a seminar on it, he really wanted to understand everything. He had no tolerance for ambiguity and weak unfounded arguments and did not hide his aversion to them. While he considered it his privilege and even duty to interrupt the speaker whenever he felt like it, he was much less patient with questions of other people (e.g. me), expecting them to think carefully before asking. *Here is another lesson I learned:* **the importance of an exciting informal weekly seminar series (or group meetings).**

Michael was a wonderful teacher and lecturer. **The pride he took in his teaching and the time he devoted to plan and prepare his lectures – classes as well as scientific seminars** *– were yet another aspect in which he served as an example for me.* At those prehistoric, pre-powerpoint times overhead projectors were used to show handwritten transparencies. Michael insisted on using two projectors simultaneously, leaving a transparency on while the next one was introduced, making the transition between them smoother and easier to follow. Clearly the two projectors had to be close enough so that he did not have to run long distance between them, but far enough to keep the projections well separated (on two screens, of course). His transparencies were meticulously prepared and he liked artistic effects and surprises. When talking about critical exponents, he compared the increasing accuracy with which they were calculated and measured with our improving knowledge of $\pi$. He heard that somewhere in the

---

[g] There were exceptions. Roald Hoffmann from the Chemistry Department was allowed transparencies, and when I asked Michael how come? He answered that people who got or will get the Nobel Prize were exempt of the no transparencies rule. I have not heard yet of the Woodward-Hoffmann rules, but sure enough, Hoffmann received the Nobel Prize some years later. Interestingly, neither Ken Wilson nor Mike Kosterlitz were exempt.

Bible the value π=3 appeared, and asked me whether I knew about this. I asked around and informed him that indeed, there is mention of a large circular copper basin cast for Solomon's Temple, "a molten sea, ten cubits from the one brim to the other: it was round all about… and a line of thirty cubits did compass it round about"[17].  He was delighted and asked me to make a copy of the original Hebrew page, which he showed proudly for years.

Returning to the opening Einstein quote, I recall one instance where Michael served for me as a "warning example". One day I told him of an idea I had – to combine the RG and Monte Carlo. He gave me a dismissive look and said: "you will combine the evil of the two methods". I did not pursue the idea, and a few months later Sheng Ma[18] published a Letter on this, followed two years afterwards by the wonderful work of Bob Swendsen.[19] As a result, whenever one of my students announced that she (or he) had an idea, no matter how far out it sounded, I always expressed surprise and pleasure and encouraged her to work on it for a few weeks – and then either drop it or come see me again about continuing. Someone having an idea is a cause for celebration! I beware of extinguishing the quest for new ideas.

One other role Michael took, which of course I could not adopt, was that of *a "super-referee" of the field. Anyone who published a paper on phase transitions and critical phenomena had to be prepared to receive a letter from Fisher, complimenting his work and expressing criticism of parts that he did not agree with.* Everyone, especially people who worked with him in the past, always had this super-referee in mind, asking ourselves – I wonder, what will Fisher think about this? His involvement gave every now and then rise to amusing incidents; I will mention two, one of which I only heard second hand, while about the second I am confident, having seen the correspondence.

The first story concerns a Letter on exchange effects in bcc $^3$He, by J.H. Hetherington and F.D.C. Willard.[20] Michael did not like something about it, assumed that Willard was the senior author, and rushed off a "Dear Prof. Willard" letter. As it turned out, it was originally a single author paper by Hetherington who used plural form all over.  When his Letter was accepted, the copy editor requested to change "we" to "I" everywhere, which meant retyping a technically difficult paper. Hetherington preferred to add an author, F. D. C. Willard,[h] his cat, who now had to correspond with Fisher!

The second story concerns S. Shtrikman, a very prominent Israeli physicist whom I knew and respected, and was surprised when Michael complained about him. When I asked him why, he sent me a letter he wrote to Shtrikman, reprimanding him about a serious oversight: identifying[21] the measured (finite) peak of the susceptibility of an antiferromagnet as the Neel temperature, whereas Michael has shown that $T_N$ was, in fact, at a temperature close but below that of the peak, where the derivative of the function diverged.[22] He was offended by Shtrikman not bothering to answer his letter. I looked at the letter and asked him – have you noticed the date? May 30, 1967. – So? I told him that this was a week before the 6 day war started and even if the letter ever arrived, Shtrikman was probably on reserve duty and had other things on his mind. Not a man who concedes easily, Michael alerted me to the fact that Hornreich, the junior co-author of the paper was at the time on Sabbatical at Yale, and could have answered.

---

[h] F. D. C. stood for Felix Domesticus Chester.

No account of Michael would be complete without mention of his family. I remember fondly the huge house in Ithaca, presided over by Sorrel, the kindest, funniest and warmest person you met. The Christmas lunch morphing into dinner, with the strange Christmas pudding, a round object made of undefined ingredients that has been buried for six months before being dug up and consumed – I have no idea whether this was true or invented to check my naivety. The house was full of annoying puzzles that were stuck in your hand with the challenge of taking them apart, putting them together, or remove them from the buttonhole of your jacket. The flamenco guitar, which had a seat of its own on flights to Israel. The Fisher Kids, of whom he was enormously proud.

### 6. Wrapping up, post Cornell

In the spring of 76 my wife informed me that "I do not know about you, but this past winter was *my* last one in Ithaca". The original plan was to stay for four years at Cornell, i.e. till Fall 77, so I told Michael that I will have to leave sooner. It was time to write a Ph D thesis, and the way this happened brought to fore *two aspects of Michael which I also tried to adopt:* ***foresight about theses and dealing with administration firmly but without confrontation.***

When I wrote my first crossover paper, the last thing I had on my mind was my thesis. Michael told me to make sure that the final version was typed precisely in the format of a Cornell thesis. The same held for the second paper. Now when the time came for writing the thesis, I already had the main part of it done and ready! I wrote a few brief introductory Chapters, and was all set to put everything together, when disaster struck. When the second paper was ready for typing, Michael's secretary Lynn was either ill or overloaded, and it was farmed out to an excellent professional typist in Ithaca, who did a superb job. Her IBM typewriter had, however, a "head" with a slightly different font than Lynn's. When I discovered this I rushed to warn Michael that the secretary in charge of quality control of theses had a reputation of being quite difficult and the rules were very explicit about using the same typewriter throughout the entire thesis. Departure date for my postdoc was already set and time was starting to run out. Michael looked with disbelief at the typescripts of the two papers and exclaimed: "but they are practically identical!" I insisted – if I noticed the difference, no doubt the secretary will also discover it. I was curious to see how will Michael fight this out, and was impressed watching him in action. He wrote an extremely polite lengthy letter to the Dean in charge, explaining to her the problem I was working on, how technically difficult it was to type this thesis, the problems with Lynn, how hard we tried to match typewriters and alas, one of the Chapters is in a barely noticeably different font. Would she be so kind to allow this, just for once? When two weeks passed and there was no answer, Michael wrote a second letter, thanking the Dean profusely for her flexibility and agreement to his request, of having one Chapter in a slightly different font. When I went, shaking, to see the secretary whose approval was essential for submission, she immediately caught the two distinct typewriters. I informed her that Professor Fisher had discussed the matter extensively with the Dean, an agreement had been reached, and produced a copy of Michael's letter of thanks. She was shaken but went on to check the thesis. Her eyes lit up when she discovered, to my horror, that in the "Chapter of the Second Typewriter" a typo, about which I completely forgot, had been corrected using another typewriter! She kicked me out, saying – the Dean may have agreed to two typewriters, but not to three! Michael phoned the lady who typed it for us, begged her to drop whatever she was doing and retype the single offensive page. I rushed over and sat with her while she did it, rushed back to the secretary, had the thesis approved, photocopied everything on paper of the right weight and submitted it a few hours before our plane for my postdoc at Seattle took off, three years and four months after my arrival at Cornell.

From teacher and student we progressed over the years to friendship. We both liked to write, so many letters were exchanged. He came a lot to Israel, as member of the Weizmann Board (see Fig. 2) and various Scientific and Academic Advisory Committees, driving Physicists, Mathematicians and Chemists crazy with his questions, critique and advice. He always spoke his mind in a clear, loud and direct fashion, but was also aware of the politics of the place. As a young just tenured Associate Professor I campaigned to stop construction of a huge solar tower, a showy, wasteful and crazy large scale engineering project driven by some donor's whim, and told Michael about my fight. He listened and asked me a single question – who is the Weizmann scientist pushing for this? I explained to him who it was and Michael summarized the situation in one sentence; "I see, so he is way out of your league", advising me to stop a futile fight. Over the years whenever I started to feel important, I walked to a window to look at the solar tower, to place things in proper perspective.

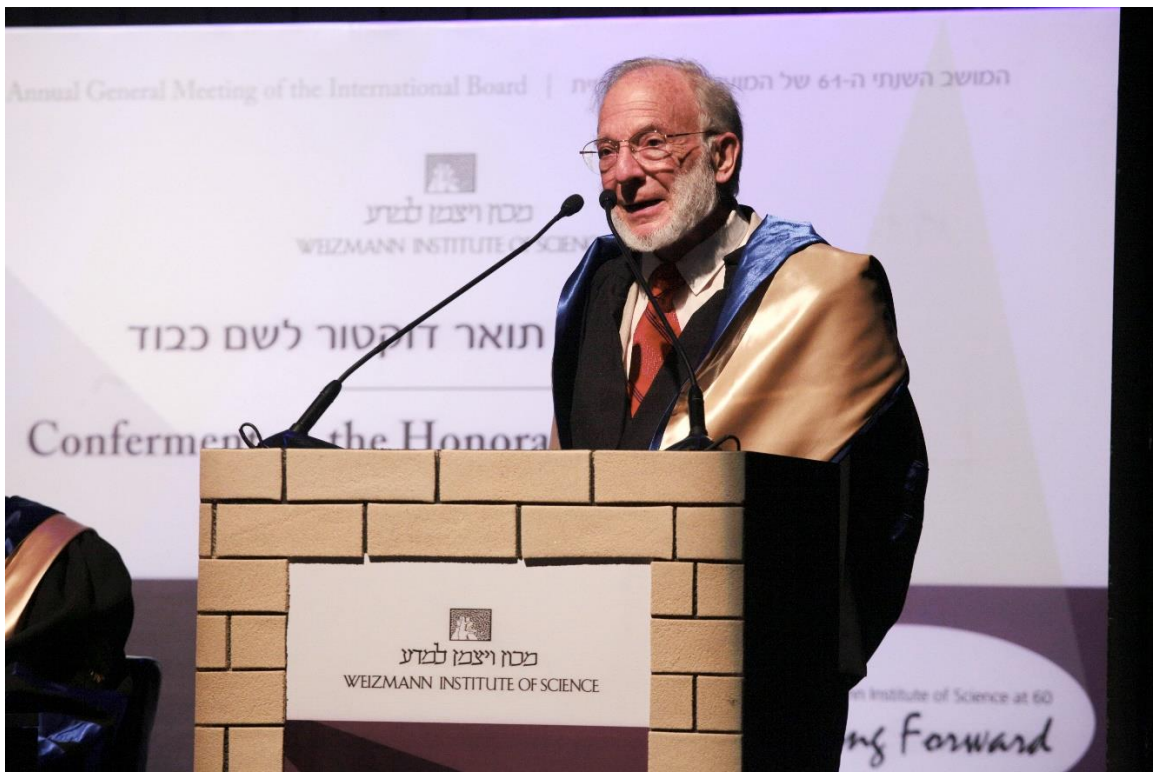

Fig. 2. Michael responding to the award of Ph D Honoris Causa from the Weizmann Institute, Nov. 2009.

After his move to Maryland I tried to pass through whenever I had a chance. I was touched when during my last visit to the fancy establishment to which Michael and Sorrel moved he gave me as a gift a Sheffield steel hunting knife that belonged to his father. The knife is on my desk, and the "gadfly and troubadour"[23] who gave it to me is carved into my heart.